\documentclass[12pt,thmsa]{article}
\usepackage{amsfonts}
\usepackage{graphicx}
\begin{document}

\title{{\Large \textbf{Quantum Computing, NP-complete Problems \\
and \\
Chaotic Dynamics}}}
\author{Masanori Ohya and Igor V. Volovich \thanks{%
Permanent address: Steklov Mathematical Institute, Gubkin St.8, GSP-1,
117966, Moscow, Russia, volovich@mi.ras.ru} \\
\\
Department of Information Sciences\\
Science University of Tokyo\\
Noda City, Chiba 278-8510, Japan\\
e-mail: ohya@is.noda.sut.ac.jp}
\date{}
\maketitle

\begin{abstract}
An approach to the solution of NP-complete problems based on quantum
computing and chaotic dynamics is proposed. We consider the satisfiability
problem and argue that the problem, in principle, can be solved in
polynomial time if we combine the quantum computer with the chaotic dynamics
amplifier based on the logistic map. We discuss a possible implementation of
such a chaotic quantum computation by using the atomic quantum computer with
quantum gates described by the Hartree-Fock equations. In this case, in
principle, one can build not only standard linear quantum gates but also
nonlinear gates and moreover they obey to Fermi statistics. This new type of
entaglement related with Fermi statistics can be interesting also for
quantum communication theory.
\end{abstract}

\newpage

\section{Introduction}

There are important problems such as the napsack problem, the traveling
salesman problem, the integer programming problem, the subgraph isomorphism
problem, the satisfiability problem that have been studied for decades and
for which all known algorithms have a running time that is exponential in
the length of the input. These five problems and many other problems belong
to the set of \textbf{NP}-complete problems. Any problem that can be solved
in polynomial time on a nondeterministic Turing machine is polynomially
transformed to an \textbf{NP}-complete problem \cite{GJ}.

Many \textbf{NP}-complete problems have been identified, and it seems that
such problems are very difficult and probably exponential. If so, solutions
are still needed, and in this paper we consider an approach to these
problems based on quantum computers and chaotic dynamics.

It is widely believed that quantum computers are more efficient than
classical computers. In particular Shor \cite{Sho} gave a remarkable quantum
polynomial-time algorithm for the factoring problem. However, it is unknown
whether this problem is \textbf{NP}-complete.

The computational power of quantum computers has been explored in a number
of papers. Bernstein and Vasirani \cite{BV} proved that \textbf{BPP}$%
\subseteq $\textbf{BQP}$\subseteq $ \textbf{PSPACE}. Here \textbf{BPP }
stands for the class of problems efficiently solvable in the classical
sense, i.e., the class of problems that can be solved in polynomial time by
probabilistic Turing machines with error probability bounded by 1/3 for all
inputs. The quantum analogue of the class \textbf{BPP }is the class \textbf{%
BQP} which is the class of languages that can be solved in polynomial time
by quantum Turing machines with error probability bounded by 1/3 for all
inputs.

The question whether \textbf{NP}$\subseteq $\textbf{BQP, }i.e., can quantum
computers solve \textbf{NP}-complete problems in polynomial time, was
considered in \cite{BBBV}. It was proved in \cite{BBBV} that relative to an
oracle chosen uniformly at random, with probability 1, the class \textbf{NP}
can not be solved on a quantum Turing machine in time o$\left(
2^{n/2}\right) .$ An oracle is a special subroutine call whose invocation
only costs unit time. This result does not rule out the possibility that 
\textbf{NP}$\subseteq $\textbf{BQP }but it does establish that there is no
black-box approach to solving \textbf{NP}-complete problems in polynomial
time on quantum Turing machines. We would like to mention that these results
are not immediately applicable to the chaotic quantum computer which we
consider in this paper.

For a recent discussion of computational complexity in quantum computing see 
\cite{FR,Cle,HHZ}. Mathematical features of quantum computing and quantum
information theory are summarized in \cite{O1}. A possibility to exploit
nonlinear quantum mechanics so that the class of problems \textbf{NP} may be
solved in polynomial time has been considered by Abrams and Lloyd in \cite
{AL}. It is mentioned in \cite{AL} that such nonlinearity is purely
hypotetical; all known experiments confirm the linearity of quantum
mechanics.

The satisfiability problem (SAT), which is \textbf{NP}-complete problem, has
been considered in quantum computing in \cite{OM}. It was shown in \cite{OM}
that the SAT problem can be solved in polynomial time by using a quantum
computer under the assumption that a special superposition of two orthogonal
vectors can be physically detected . The problem one has to overcome here is
that the output of computations could be a very small number and one needs
to amplify it to a reasonable large quantity.

In this paper we propose that chaotic dynamics plays a constructive role in
computations. Chaos and quantum decoherence are considered normally as the
degrading effects which lead to an unwelcome increase of the error rate with
the input size. However, in this paper we argue that under some
circumstances chaos can play a constructive role in computer science. In
particular we propose to combine quantum computer with the chaotic dynamics
amplifier. We will argue, by using the consideration from \cite{OM}, that
such a chaotic quantum computer can solve the SAT problem in polynomial time.

As a possible specific implementation of chaotic quantum computations we
discuss the recently proposed atomic quantum computer \cite{Vol1}. It is
proposed in \cite{Vol1} to use a \textit{single }atom as a quantum computer.
One can implement a single qubit in atom as a one-particle electron state in
the self-consistent field approximation and multi-qubit states as the
corresponding multi-electron states represented by the Slater determinant.

A possible realization of the standard quantum gates in the atomic quantum
computer by using the electron spin resonance has been discussed in \cite
{Vol1}. In this paper we argue that in the atomic quantum computer one can
build also \textit{nonlinear} quantum gates because the dynamics of the
multi-electron atom in the very good approximation is described by nonlinear
Hartree-Fock equations.

The tensor product structure of states is very important for computations
and the multielectron atom admits such a structure. More exactly, instead of
the standard tensor product used in quantum computing we have to use the
Slater determinant to take into account the Fermi statistics.The standard
computational basis in quantum computing does not have Bose or Fermi
symmetry. In the atomic case we have to make an appropriate modification of
quantum gates to take into account Fermi statistics and this leads to a new
type of entanglement related with Fermi statistics.

Such Fermi or Bose entanglement could be interesting also for quantum
communication theory, in particular for quantum teleportation
\cite{AO,FO}.

\section{SAT Problem}

Let $\left\{ x_{1},\cdots ,x_{n}\right\} $ be a set of Boolean variables, $%
x_{i}=0$ or $1.$ Then the set of the Boolean variables $\left\{ x_{1}, 
\overline{x}_{1},\cdots ,x_{n},\overline{x}_{n}\right\} $ with or without
complementation is called the set of \textit{literals.} A formula, which is
the product (AND) of disjunctions (OR) of literals is said to be in the 
\textit{product of sums }(POS) form. For example, the formula 
\[
\left( x_{1}\vee \overline{x}_{2}\right) \left( \overline{x}_{1}\right)
\left( x_{2}\vee \overline{x}_{3}\right) 
\]
is in POS form. The disjunctions $\left( x_{1}\vee \overline{x}_{2}\right)
,\left( \overline{x}_{1}\right) ,\left( x_{2}\vee \overline{x}_{3}\right) $
here are called \textit{clauses.} A formula in POS form is said to be 
\textit{satisfiable }if there is an assignment of values to variables so
that the formula has value 1. The preceding formula is satisfiable when $%
x_{1}=0,$ $x_{2}=0,$ $x_{3}=0.$

\textbf{Definition}(SAT Problem). The satisfiability problem (SAT) is to
determine whether or not a formula in POS form is satisfiable.

The following analytical formulation of SAT problem is useful. We define a
family of Boolean polynomials $f_{\alpha }$, indexed by the following data.
One $\alpha $ is a set 
\[
\alpha =\left\{ S_{1},...,S_{N},T_{1},...,T_{N}\right\} , 
\]
where $S_{i},T_{i}\subseteq \left\{ 1,...,n\right\} ,$ and $f_{\alpha }$ is
defined as 
\[
f_{\alpha }(x_{1},\cdots ,x_{n})=\prod_{i=1}^{N}\left( 1+\prod_{a\in
S_{i}}(1+x_{a})\prod_{b\in T_{i}}x_{b}\right) . 
\]

We assume here the addition modulo 2. The SAT problem now is to determine
whether or not there exists a value of $\mathbf{x}=(x_{1},\cdots ,x_{n})$
such that $f_{\alpha }(\mathbf{x})=1.$

\section{Quantum Algorithm}

We will work in the $\left( n+1\right) $-tuple tensor product Hilbert space $%
\mathcal{H\equiv }$ $\otimes _{1}^{n+1}$\textbf{C}$^{2}$ with the
computational basis 
\[
\left| x_{1},...,x_{n},y\right\rangle =\otimes _{i=1}^{n}\left|
x_{i}\right\rangle \otimes \left| y\right\rangle 
\]
where $x_{1},...,x_{n},$ $y=0$ or $1.$ We denote $\left|
x_{1},...,x_{n},y\right\rangle =\left| \mathbf{x},y\right\rangle .$ The
quantum version of the function $f(\mathbf{x})=f_{\alpha }(\mathbf{x})$ is
given by the unitary operator $U_{f}\left| \mathbf{x},y\right\rangle =\left| 
\mathbf{x},y+f(\mathbf{x})\right\rangle .$ We assume that the unitary matrix 
$U_{f}$ can be build in the polynomial time, see \cite{OM}. Now let us use
the usual quantum algorithm:

(i) By using the Fourier transform produce from $\left| \mathbf{0,}
0\right\rangle $ the superposition 
\[
\left| v\right\rangle =\frac{1}{\sqrt{2^{n}}} \sum_{\mathbf{x}}\left| 
\mathbf{x},0\right\rangle . 
\]

(ii) Use the unitary matrix $U_{f}$ to calculate $f(\mathbf{x}):$%
\[
\left| v_{f}\right\rangle =U_{f}\left| v\right\rangle =\frac{1}{\sqrt{2^{n}}}
\sum_{\mathbf{x}}\left| \mathbf{x},f(\mathbf{x})\right\rangle 
\]
Now if we measure the last qubit, i.e., apply the projector $P=I\otimes
\left| 1\right\rangle \left\langle 1\right| $ to the state $\left|
v_{f}\right\rangle ,$ then we obtain that the probability to find the result 
$f(\mathbf{x})=1$ is $\left\| P\left| v_{f}\right\rangle \right\|
^{2}=r/2^{n}$ where $r$ is the number of roots of the equation $f(\mathbf{x}
)=1.$ For small $r $ the probability is very small and this means we in fact
don't get an information about the existence of the solution of the equation 
$f(\mathbf{x} )=1.$ Let us simplify our notations. After the step (ii) the
quantum computer will be in the state 
\[
\left| v_{f}\right\rangle =\sqrt{1-q^{2}}\left| \varphi _{0}\right\rangle
\otimes \left| 0\right\rangle +q\left| \varphi _{1}\right\rangle \otimes
\left| 1\right\rangle 
\]
where $\left| \varphi _{1}\right\rangle $ and $\left| \varphi
_{0}\right\rangle $ are normalized $n$ qubit states and 
$q=\sqrt{r/2^{n}}.$
Effectively our problem is reduced to the following $1$ qubit problem. We
have the state 
\[
\left| \psi \right\rangle =\sqrt{1-q^{2}}\left| 0\right\rangle +q\left|
1\right\rangle 
\]
and we want to distinguish between the cases $q=0$ 
(i.e. very small $q$) and $q>0$. To this end we
propose to employ chaotic dynamics.

\section{Chaotic Dynamics}

Various aspects of classical and quantum chaos have been the subject of
numerious studies, see \cite{O2} and ref's therein.The investigation of
quantum chaos by using quantum computers has been proposed in \cite{
GCZ,Sch,KM}. Here we will argue that chaos can play a constructive role in
computations.

Chaotic behaviour in a classical system usually is considered as an
exponential sensitivity to initial conditions. It is this sensitivity we
would like to use to distinquish between the cases $q=0$ and $q>0$ from the
previous section.

Consider the so called logistic map which is given by the equation 
\[
x_{n+1}=ax_{n}(1-x_{n}),~~~x_{n}\in \left[ 0,1\right] .
\]

\noindent \noindent \noindent The properties of the map depend on the
parameter $a.$ If we take, for example, $a=3.71,$ then the Lyapunov exponent
is positive, the trajectory is very sensitive to the initial value and one
has the chaotic behaviour \cite{O2}. It is important to notice that if the
initial value $x_{0}=0,$ then $x_{n}=0$ for all $n.$

\begin{center}
\includegraphics{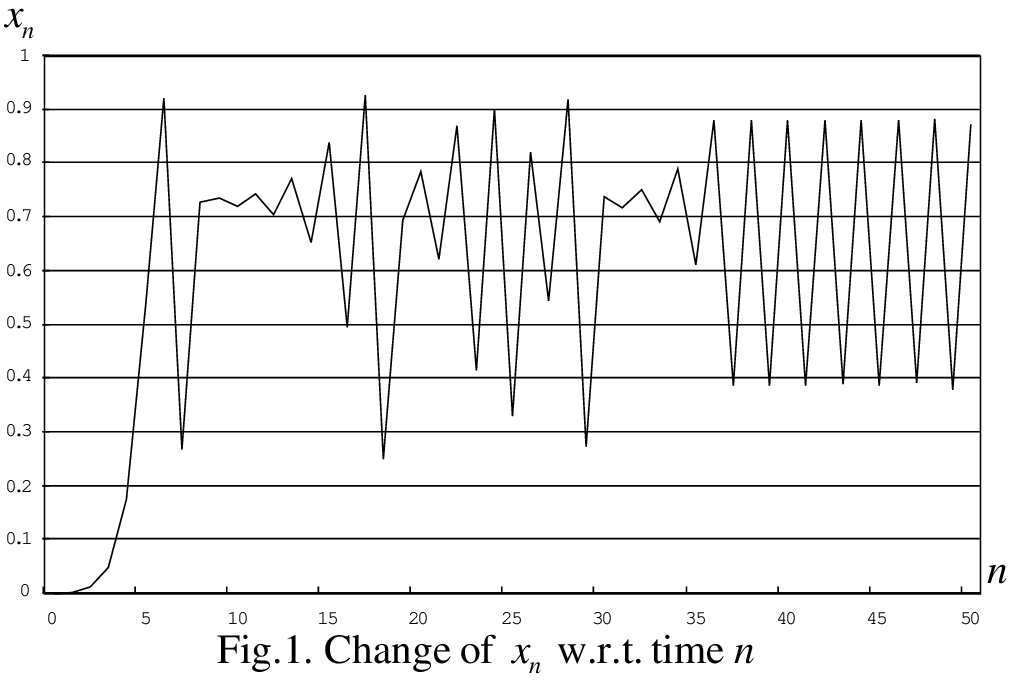}
\end{center}

It is known \cite{Deu} that any classical algorithm can be implemented on
quantum computer. Our stochastic quantum computer will be consisting from
two blocks. One block is the ordinary quantum computer performing
computations with the output $\left| \psi \right\rangle =\sqrt{1-q^{2}}%
\left| 0\right\rangle +q\left| 1\right\rangle $. The second block is a 
\textit{quantum} computer performing computations of the \textit{classical}
logistic map. This two blocks should be connected in such a way that the
state $\left| \psi \right\rangle $ first be transformed into the density
matrix of the form 
\[
\rho =q^{2}P_{1}+\left( 1-q^{2}\right) P_{0} 
\]
where $P_{1}$ and $P_{0}$ are projectors to the states $\left|
1\right\rangle $ and $\left| 0\right\rangle .$ This connection is in fact
nontrivial and actually it should be considered as the third block. One has
to notice that $P_{1}$ and $P_{0}$ generate an Abelian algebra which can be
considered as a classical system. In the second block the density matrix $%
\rho $ above is interpreted as the initial data $\rho _{0}$ for the logistic
map 
\[
\rho _{n+1}=a\rho _{n}(1-\rho _{n}) 
\]
After one step, the state $\rho _{1}$ becomes 
\[
\rho _{1}=aq^{2}(1-q^{2})I, 
\]
where $I$ is the identity matrix on $\Bbb{C}^{2}.$ In paricular, if one has $%
q=0$ then $\rho _{0}=P_{0}$ and we obtain $\rho _{n}=P_{0}$ for all $n.$
Otherwise the stochastic dynamics leads to the amplification of the small
magnitude $q$ in such a way that it can be detected. As is seen in Fig.1, we
can easily amplify the small $q$ in several steps, i.e., within about ten
times measurements as in Shor's algorithm. 
The transition from $\rho _{n}$ to $%
\rho _{n+1}$ is nonlinear and can be considered as a discrete Heisenberg
evolution of the variable $x_{n}$.

One can think about various possible implementations of the idea of using
chaotic dynamics for computations. Below we discuss how one can realize
nonlinear quantum gates on atomic quantum computer.

\section{\noindent Atomic Quantum Computer}

Many current proposals for the realization of quantum computer such as NMR,
quantum dots and trapped ions are based on the using of an atom or an ion as
one qubit, see \cite{CZ,GC,BLD,EJ}. In these proposals a quantum computer
consists from several atoms, and the coupling between them provides the
coupling between qubits necessary for a quantum gate. It was proposed in 
\cite{Vol1} that a \textit{single} atom can be used as a quantum computer.
One can implement a single qubit in atom as a one-particle electron state in
the self-consistent field approximation and multi-qubit states as the
corresponding multi-electron states represented by the Slater determinant.
So, to represent 10 qubits one can use an atom with 10 electrons and to
represent 50 qubits one has to control only around 50 levels in an atom with
50 electrons.

A possible realization of the standard quantum gates in the atomic quantum
computer by using the electron spin resonance has been discussed in \cite
{Vol1}. In this paper we propose that in the atomic quantum computer one can
build also \textit{nonlinear} quantum gates because the dynamics of the
multi-electron atom in the very good approximation is described by nonlinear
Hartree-Fock equations. Therefore it follows from \cite{OM} and the
considerations in this paper that the atomic quantum computer can solve the
SAT problem in polynomial time.

It is well known that in atomic physics the concept of the individual state
of an electron in an atom is accepted and one proceeds from the Hartree-Fock
self-consistent field approximation, see for example \cite{Sob}. The state
of an atom is determined by the set of the states of the electrons. Each
state of the electron is characterized by a definite value of its orbital
angular momentum $l$, by the principal quantum number $n$ and by the values
of the projections of the orbital angular momentum $m_{l}$ and of the spin $%
m_{s}$ on the $z$-axis. In the Hartree-Fock central field approximation the
energy of an atom is completely determined by the assignment of the electron
configuration, i.e., by the assignment of the values of $n$ and $l$ for all
the electrons.

The tensor product structure of states is very important for computations.
Fortunately a multielectron atom admits such a structure. More exactly,
instead of the standard tensor product used in quantum computing we have to
use the Slater determinant to take into account the Fermi statistics.The
standard computational basis in quantum computing does not have Bose or
Fermi symmetry. In the atomic case we have to make an appropriate
modification of quantum gates to take into account Fermi statistics and this
leads to a new type of entanglement related with Fermi statistics.

An application of the electron spin resonance (ESR) to process the
information encoded in the hyperfine splitting of atomic energy levels and
to build standard linear quantum gates has been considered in \cite{Vol1}.
In this paper we suggest that in atomic quantum computer one can build also 
\textit{nonlinear} quantum gates described by the Hartree-Fock equations.

The Hamiltonian for the $N-$ particle system has the form 
\[
H=\sum_{i=1}^{N}\left( -\frac{\nabla _{i}^{2}}{2m_{i}}+v(r_{i})\right)
+\sum_{i<j}V(r_{i},r_{j}) 
\]
In the Hartree-Fock method one takes the $N-$ particle wave function in the
form of the Slater determinant 
\[
\Psi (t,r_{1},...,r_{N})=Antisym(\Phi _{1}(t,r_{1})...\Phi _{N}(t,r_{N})) 
\]
Here the one-particle wave functions $\Phi _{i}(t,r_{i})$ satisfy the
nonlinear Hartree-Fock equations which have the form of nonlinear
Schrodinger equation 
\[
i\frac{\partial \Phi _{i}(t,r)}{\partial t}=\mathit{H}(\Phi )\Phi _{i}(t,r) 
\]
where 
\[
\mathit{H}(\Phi )\Phi _{i}(t,r)=\left( -\frac{\nabla ^{2}}{2m_{i}}+v(r)+%
\mathcal{U}_i(t,r)\right) \Phi _{i}(t,r)-\int dr^{\prime }\mathcal{W}%
(t,r^{\prime },r)\Phi _{i}(t,r^{\prime }) 
\]
and 
\[
\mathcal{U}_i\left( t,r\right) =\sum_{j\neq i}\int dr^{\prime }\Phi
_{j}^{*}\left( t,r^{\prime }\right) V\left( r^{\prime },r\right) \Phi
_{j}\left( t,r^{\prime }\right) , 
\]

\[
\mathcal{W}\left( t,r^{\prime },r\right) = \sum_{j}\Phi _{j}^{*}\left(
t,r^{\prime }\right) V\left( r^{\prime },r\right) \Phi _{j}\left(
t,r\right) 
\]

If we consider only the spin dependent part of the wave function of the
one-electron state 
\[
\varphi =\left( 
\begin{array}{c}
\varphi _{0}\left( t\right) \\ 
\varphi _{1}\left( t\right)
\end{array}
\right) , 
\]
then one can get the nonlinear equation of the form 
\[
i\frac{\partial \varphi }{\partial t}=A\varphi +B\left( \varphi \right)
\varphi . 
\]
Here $A$ is a $2\times 2$ matrix and the matrix $B$ depends on $\varphi .$
By using this equation one can describe nonlinear quantum gate. The
nonlinearity can be tuned by means of magnetic field.

\section{Conclusion}

The complexity of the quantum algprithm for the SAT problem has been
considered in \cite{OM} where it was shown that one can build the unitary
matrix $U_{f}$ in the polynomial time. We have also to consider the number
of steps in the classical algorithm for the logistic map performed on
quantum computer. It is the probabilistic part of the construction and one
has to repeat computations several times to be able to distingish the cases $%
q=0$ and $q>0.$ Thus it seems that the chaotic quantum computer can solve
the SAT problem in polynominal time. 

In conclusion, in this paper the
chaotic quantum computer is proposed. It combines the ordinary quantum
computer with quantum chaotic dynamics amplifier which can be implemented by
using the atomic quantum computer. We argued that such a device can be
powerful enough to solve the \textbf{NP}-complete problem 
in polynomial time.


\begin{thebibliography}{99}
\bibitem{GJ}  M. Garey and D. Johnson, \textit{Computers and Intractability
- a guide to the theory of NP-completeness}, Freeman, 1979.

\bibitem{Sho}  P.W. Shor, \textit{Algorithm for quantum computation :
Discrete logarithm and factoring algorithm}, Proceedings of the 35th Annual
IEEE Symposium on Foundation of Computer Science, pp.124-134, 1994.

\bibitem{BV}  E. Bernstein and U. Vazirani, \textit{Quantum Complexity Theory%
}, in: Proc. of the 25th Annual ACM Symposium on Theory of Comuting, (ACM
Press, New York,1993), pp.11-20.

\bibitem{BBBV}  C. H. Bennett, E. Bernstein, G. Brassard, U. Vazirani,%
\newline
\textit{Strengths and Weaknesses of Quantum Computing}, quant-ph/9701001

\bibitem{FR}  L. Fortnow and J. Rogers, \textit{Complexity Limitations on
Quantum Computation}, cs.CC/9811023.

\bibitem{Cle}  R. Cleve, \textit{An Introduction to Quantum 
Complexity Theory%
}, quant-ph/9906111.

\bibitem{HHZ}  E. Hemaspaandra, L.A. Hemaspaandra and M. Zimand,\newline
\textit{Almost-Everywhere Superiority for Quantum Polynomial Time},
quant-ph/9910033.

\bibitem{O1}  M. Ohya, \textit{Mathematical Foundation of Quantum Computer,}
Maruzen Publ. Company, 1998

\bibitem{AL}  D. S. Abrams and S. Lloyd, \textit{Nonlinear quantum mechanics
implies polynomial-time solution for NP-complete and \#P problems,}
quant-ph/9801041.

\bibitem{OM}  M. Ohya and N. Masuda, \textit{NP problem in Quantum Algorithm,%
} quant-ph/9809075.

\bibitem{Vol1}  I.V. Volovich, \textit{Atomic Quantum Computer,}
quant-ph/9911062.

\bibitem{O2}  M. Ohya, \textit{Complexities and Their Applications to
Characterization of Chaos,} Int. Journ. of Theoret. Physics, 37 (1998) 495.

\bibitem{GCZ}  S.A. Gardiner, J.I. Cirac and P. Zoller, Phys. Rev. Lett.
79(1997) 4790.

\bibitem{Sch}  R. Schack, Phys. Rev. A57 (1998) 1634; T. Brun and R. Schack,
quant-ph/9807050.

\bibitem{KM}  I. Kim and G. Mahler, \textit{Quantum Chaos in Quantum Turing
Machine,} quant-ph/9910068.

\bibitem{Deu}  D. Deutsch, \textit{Quantum theory, the Church-Turing
principle and the universal quantum computer,} Proc. of Royal Society of
London series A, \textbf{400}, pp.97-117, 1985.

\bibitem{CZ}  J.I. Cirac and P. Zoller, Phys. Rev. Lett., 74 (1995) 74.

\bibitem{GC}  N.A. Gershenfeld and I.L. Chuang, Science, 275 (1997) 350.

\bibitem{BLD}  G. Burkard, D. Loss and D.P. DiVincenzo, cond-mat/9808026.

\bibitem{EJ}  A. Ekert and R. Jozsa, \textit{Quantum computation and Shor's
factoring algorithm,} Reviews of Modern Physics, \textbf{68}
No.3,pp.733-753, 1996.

\bibitem{Sob}  I.I. Sobelman, \textit{Atomic Spectra and Radiative
Transitions,} Springer-Verlag, 1991.

\bibitem{AO}  Accardi, L. and Ohya, M.: \textit{Teleportation of
general quantum states,} quant-ph/9912087.

\bibitem{FO}  Fichtner, K.-H. and Ohya, M.:\textit{Quantum Teleportation
with Entangled States given by Beam Splittings, } quant-ph/9912083.
\end{thebibliography}
\end{document}